\documentclass[referee]{aa} 
\usepackage{graphicx}
\usepackage{txfonts}
\usepackage{aalongtable}
\begin{document}
   \title{Variability of Optical \ion{Fe}{ii} Complex in Narrow-Line Seyfert 1 Galaxy NGC\,4051}

   \author{J. Wang
          \inst{1}
          \and
           J. Y. Wei\inst{2}
          \and
          X. T. He\inst{3}
          }

   \offprints{J. Wang}

    \institute{National Astronomical Observatories, Chinese Academy of Sciences, Beijing 100012, China\\
              \email{wj@bao.ac.cn}
         \and
              National Astronomical Observatories, Chinese Academy of Sciences, Beijing 100012, China\\
              \email{wjy@bao.ac.cn}
         \and
             Department of Astronomy, Beijing Normal University, Beijing 100875, China\\
              \email{xthe@bun.edu.cn}
             }

   \date{Received September 15, 2004; accepted  January 24, 2005}

   \abstract{

      The variability of optical \ion{Fe}{ii} blends
      in NGC\,4051 is examined from spectra extracted from the AGN Watch 
      program.  
      In our analysis, the \ion{Fe}{ii} complex are subtracted
      and measured with the following results. Firstly, 
      the \ion{Fe}{ii} variations
      were detected in NGC\,4051 during a 3-year period. 
      The identified \ion{Fe}{ii} variations  
      followed the variations in continuum closely. 
      Secondly, the EW of \ion{Fe}{ii} is reported to increase with 
      the rising continuum flux, which is consistent with previous 
      claims that there is no convincing Baldwin Effect 
      in optical \ion{Fe}{ii}. Thirdly, 
      by comparing the variations of H$\beta$ and \ion{Fe}{ii}, we
      find that $\rm{R_{Fe}}$ scales with continuum flux 
      as $\rm{R_{Fe}\propto (5.0\pm0.8) \log(L/M)}$, which is 
      significantly different from the theoretical expectations. 
      Finally, in six selected Seyferts, four out of five Narrow-Line Seyfert 1 galaxies
      present positive correlation between $\rm{R_{Fe}}$ and the continuum flux.
      The negative correlations are identified in the remaindng two
      objects that have relatively broad profiles of H$\beta$ ($\rm{FWHM>1500\ km\ s^{-1}}$).
      We argue that the difference of electron density of broad line clouds and/or variability 
      behavior of incident high-energy radiation can explain the dichotomy in 
      variability behavior of $\rm{R_{Fe}}$.

   }
   \titlerunning{Variability of \ion{Fe}{ii} Complex in NGC\,4051}
   \maketitle
%

\section{Introduction}

   Generally, the optical/UV spectra of Active Galactic Nuclei (AGNs) are prominently featured by 
   broad and intense emission lines. The relative strength and radiative mechanism of 
   broad emission lines can be interpreted well by the traditional
   photoionization models of many dense ($\sim 10^{9-10}\mathrm{cm}^{-3}$) clouds.
   Those clouds with approximate cosmic abundance are photoionized and heated by a covered 
   central source (e.g. Davidson \& Netzer 1979, Kwan \& Krolik 1981, 
   Kwan 1984, Stasinska 1984, Collin-Souffrin et al. 1988, Korista et al. 1997).
   Despite the many successes of the standard photoionization models, important 
   problems still remain. 

   Although collisional ionization is carefully introduced,
   the photoionization models are frustrated in the aspect that 
   accounts for the strength of \ion{Fe}{ii} emission in both UV and optical bands
   (Netzer \& Wills 1983, Wills et al. 1985, Joly 1987,
   Collin-Souffrin et al. 1986,1988, Dultzin-Hacyan 1987). If the total flux of \ion{Fe}{ii} is several
   times larger than Ly$\alpha$ (Wills et al. 1985), and if its energy is derived from photoionization,
   the models indicate that 
   the X-ray flux must then be comparable to or greater than the Lyman continuum flux. 
   This seems unlikely,however, given our present understanding.  
   Early calculations of \ion{Fe}{ii} emission predicted that the flux ratio of total
   \ion{Fe}{ii} to H$\beta$ was no more than 8. However, the observed \ion{Fe}{ii}/H$\beta$ is more
   typical by 12 and approaches about 30 in a few sources with ``super-strong''
   \ion{Fe}{ii} emission (Lawrence et al. 1988, Graham et al.1996, Lipari 1994, Moran et al. 1996, 
   Xia et al. 1999, Zhou et al. 2002). In order to interpret
   the intense \ion{Fe}{ii} emission, a number of additional excitation mechanisms have been proposed.
   Sigut \& Pradhan (1998) suggested
   that the Ly$\alpha$ fluorescent excitation of \ion{Fe}{ii} could double \ion{Fe}{ii}
   strength in both UV and optical bands effectively. Joly(1991 and references therein) put forward the 
   idea that 
   collisional excitation is the most likely process in AGN, however, did not include virtual calculations.

   The \ion{Fe}{ii} emission in AGN has consistently attracted a great deal of attention, both
   because of the problem described above and because of its importance in investigating the underlying
   physics that drives the Eigenvector 1 (E1).
   The E1 introduced by principal component analysis (PCA, Boroson \& Green 1992, hereafter BG92) contains 
   a strong anti-correlation between FWHM of H$\beta$ 
   and the flux ratio of the optical \ion{Fe}{ii} complex to the broad component of H$\beta$. 
   At the present time, it is believed that the E1 plays a vital role  
   in research of AGN phenomena (e.g. Boroson 2002, Sulentic et al. 2000a,b
   , Marziani et al. 2001, Zamanov \& Marziani 2002, Marziani et al. 2003a).
   Both observational and theoretical studies indicate that the E1 is most likely driven 
   by some elementary parameters of AGN, such as the central black hole mass, the Eddington 
   ratio, and even the orientation of accretion axis (Jarvis \& McLure 2002). 
   Investigation of the complex \ion{Fe}{ii}
   emission in AGN is, therefore, essential for further understanding E1 correlations and whole AGN phenomenon.

   Actually, so far the variability behavior
   of the \ion{Fe}{ii} complex in Seyfert galaxies has been poorly understood. In Mark\,110,
   Kollatschny et al. (2001) reported that the permitted optical \ion{Fe}{ii} complex remained constant
   within 10\% error over 10 years, 
   while the forbidden [\ion{Fe}{x}]$\lambda$6375 line was variable. Similarly, in the Seyfert 1 galaxy
   NGC\,5548 no significant variations of the optical \ion{Fe}{ii} blends (less than 20\%) were detected
   (Dietrich et al. 1993).
   On the contrary, the opposite result was
   reported in a long term optical variability watch program on Seyfert 1 galaxy
   NGC\,7603 over a period of nearly 20 years (Kollatschny et al. 2000). This object displayed remarkable variability 
   in the \ion{Fe}{ii} blends, with amplitudes on 
   same order as for the H$\alpha$ and \ion{He}{i} lines.
   Giannuzzo \& Stirpe(1996) found that, out of 12 Narrow-Line Seyfert
   1 galaxies(NLS1s), at least 4 of them presented a significant variability of 
   the {\ion{Fe}{ii}} complex with percentage variations larger than 30\%. In addition,
   considerable variations of the \ion{Fe}{ii} complex (larger than 50\%) were
   reported in the other two Seyfert 1 galaxies Akn\,120 and Fairall\,9 (Kollatschny et al. 1981,
   Kollatschny \& Fricke 1985). It is likely that further individual investigation, as well as
   subsequently statistical analyses, whould be essential to comprehend the physics governing 
   the \ion{Fe}{ii} emission in AGN.

   NGC\,4051, a well-known NLS1 natured by both narrow permitted emission lines
   ($\rm{FWHM} \approx 1100\ \rm{km\ s^{-1}}$) and a low ratio for [\ion{O}{iii}]/H$\beta$
   (Osterbrock \& Pogge 1985), was monitored for three years, from January 1996 to July 1998,
   as part of the AGN Watch campaign (Peterson et al. 2000). Peterson and his colleagues reported the 
   strong variabilities of intensity in H$\beta$ and \ion{He}{ii}$\lambda$4686. 
   Variations of the H$\beta$ line were found to lag behind the variations in continuum by $6\pm2-3$ days. 
   The time delay associated with the 
   Doppler width of H$\beta$ yields a viral mass estimation of $\sim \rm{1.1\times 10^{6}} M_\odot$ for the
   central black hole.
   They also found that the \ion{Fe}{ii} blends varied little, if at all, in the period spanning 
   three years according to the RMS spectra.

   In this paper, we investigate the variability of the optical \ion{Fe}{ii} complex in NGC\,4051
   by performing a new spectral analysis on the AGN Watch archival spectra.
   Besides investigation of this individual object, a comparison 
   of the variability behavior of \ion{Fe}{ii} between diverse AGNs is performed to reveal 
   the underlying physics governing \ion{Fe}{ii} emission.

   The paper is organized as follows. In \S2, we describe the spectral measurements in detail. 
   The analysis and immediate results are given in \S3. In \S4, we propose two possible explanations,
   and finally discuss the underlying implications.


\section{Spectra measurements}

   The archival spectra in one-dimensional FITS format were downloaded from the AGN Watch Web Site: 
   http://www-astronomy.mps.ohio-state.edu/\~\,agnwatch/. This archive contains a total of 123 spectra 
   divided into two data sets (A and B). Generally, for the wavelength coverage, Set B spectra extend shortward 
   further than do spectra of Set A.
   In our analysis, we discard the first spectrum observed at Julian Date 2,450,095.0 
   because of the unavailable FITS file. The spectra n00229b and n00600b are obviated in measurements because of their 
   poor spectral quality (i.e. bad S/N for continuum, as well as for emission lines). 
   Spectrum n00479a is excluded from spectral analysis, as well, because of its improper wavelength coverage
   only ranging between 4720\AA\ and 5990\AA.
   Detailed information on the archival spectra can be consulted in Table 1 
   and Section 2.2 in Peterson et al. (2000). In Table. 1, the file name of each of the remaining 120 spectra is 
   given in Col. (1), and Col. (2) lists the corresponding Julian Date of observation.

  \subsection{Pretreatments and \ion{Fe}{ii} measurements}

  The remaining 120 spectra are pretreated in the  following two steps: 1) the Galactic extinction is corrected
  by $E(B-V)=0.013$ mag from NED, adopting an $R_V=3.1$ extinction curve; 2) each spectrum is transformed
  to rest frame, as well as K-corrected by the redshift derived by a Gaussian fitting for the narrow peak of 
  the H$\beta$ line.
  As an illustration, the rest frame spectrum of NGC\,4051 taken on June 16, 1996 (JD = 2,450,250.7)
  is shown in Fig. 1. The spectrum covers the wavelength range from 3600\AA\ to 7540\AA, and the 
  strongest emission lines and prominent \ion{Fe}{ii} complex are labeled.

  As shown in Fig. 1, there is a clear contribution from blends of \ion{Fe}{ii} line emission on both the 
  red and blue sides of the H$\beta$-[\ion{O}{iii}] complex. The blends contaminate strong emission lines and 
  alter the fluxes of H$\beta$ and [\ion{O}{iii}]$\lambda\lambda$4959, 5007. 
  In order to determine contributions of the \ion{Fe}{ii} blends and to measure the other emission lines reliably,
  we subtract the \ion{Fe}{ii} multiplets from each observed spectrum by the experiential technique
  described by BG92. The subtraction depends on a template of the \ion{Fe}{ii} complex. In this paper,
  the adopted template is the same as that used in BG92, namely, the \ion{Fe}{ii} emission of I\,ZW1, which is a well-known
  prototype of bright NLS1 with narrow permitted \ion{Fe}{ii} emission lines (Phillips 1978, Oke \& Lauer 1979).
  The detailed procedure for making the template of the \ion{Fe}{ii} emission can be found in BG92, But, briefly,
  the template is a two-dimensional function of FWHM and intensity of the \ion{Fe}{ii} blends.
  The template can be broadened to the FWHM of the broad component of H$\beta$ by convolving with a Gaussian profile
  and scaled to match the \ion{Fe}{ii} strength.
  In NGC\,4051, the FWHM of the template
  is taken to be $\rm{1200\ km\ s^{-1}}$ which approximates the value of FWHM
  of H$\beta$ ($V_{\rm{FWHM}} = 1110 \pm 190\ \rm{km\ s^{-1}}$) derived by Peterson et al. (2000). For each spectrum,
  the scaling factor is estimated at rest wavelength 4570\AA; then the convolved and scaled templates are
  subtracted from the observed spectra. A successful \ion{Fe}{ii} subtraction requires a smooth
  continuum at blueward of H$\beta$ and between 5100\AA\ and 5500\AA.
  The \ion{Fe}{ii} subtraction is sketched in Fig. 2 for the case observed at JD = 2,450,250.7.
  In the figure, the bottom curve is the best estimated \ion{Fe}{ii} multiplets; and the \ion{Fe}{ii} subtracted 
  spectrum is shown in the middle; while the observed spectrum is plotted at the top.
  Note that the observed 
  spectrum is offset upwards arbitrarily for visibility. The errorbar of \ion{Fe}{ii} strength, 
  however, is very hard to obtained accurately because the subtraction is done by eye.
  Therefore, the uncertainties of \ion{Fe}{ii} intensities, which generally should be round about 30-50
  percent, are superseded by the upper and lower limits. These limits are carefully obtained by
  iterative experimentations with a series of values for the flux of the template.  
  Outside of the limits, the \ion{Fe}{ii} subtracted continuum is absolutely unacceptable. 

  A new \ion{Fe}{ii} template was recently published by V\'{e}ron-Cetty et al. (2004). We perform the \ion{Fe}{ii} 
  subtraction described above for 7 typical spectra in the terms of this new template. The inferred \ion{Fe}{ii} intensities
  are compared to the intensities provided by BG92's template. The two \ion{Fe}{ii} flux measurements are highly 
  correlated and consistent with a nearly linear relation. This relation indicates that both templates can archieve the 
  same result when we focus attention on the variability behavior of the optical \ion{Fe}{ii} complex.
  In addition to this relationship, the \ion{Fe}{ii} flux obtained 
  by the template of V\'{e}ron is systematically lower than the flux obtained by employing the BG92 template.
  For instance, in the spectrum observed at JD=2,450,250.7, the calibrated \ion{Fe}{ii} flux(see Sect. 3) is
  $5.25\times10^{-13}\ \rm{erg\ s^{-1}\ cm^{-2}}$ for the V\'{e}ron's template, 
  but $7.54\times10^{-13}\ \rm{erg\ s^{-1}\ cm^{-2}}$ for the BG92 template. 
  This discrepancy is quite rational because the two templates differ in their 
  narrow permitted and forbidden \ion{Fe}{ii} lines. These lines are excluded from the V\'{e}ron's template and are of 
  not negligible fluxes.

  \subsection{Line profile modelling}

  The \ion{Fe}{ii} multiplets contamination-removed spectra are characterized not only by the prominent 
  H$\beta$ and [\ion{O}{iii}]$\lambda\lambda$4959,5007 lines, but also by their broad and strong \ion{He}{ii}$\lambda$4686 
  emission. The next step in processing the spectra is to remove the continuum from each spectrum.
  Generally, the continuum is carefully modelled by a power law based upon two selected wavelength regions 4400\AA-4450\AA\
  and 5100\AA-5500\AA\ in most situations. 
  Both regions are free of any strong emission lines. The IRAF-SPECFIT task, a multi-component profile
  modelling procedure described by Kriss (1994), was utilized to model the isolated emission lines.
  V\'{e}ron-Cetty et al. (2001) claimed that it is better to 
  model H$\beta$ emission line by a Lorentzian profile than by a Gaussian profile in NLS1 galaxies. 
  The choice of 
  profile to represent the observed emission line, however, may have no physical significance(e.g. Evans 1988, Xu et al. 2003),
  especially when we focus on an integrated line flux. The following Gaussian
  components and specific relationships are involved when 
  modelling each of the spectra. For the first step, the profile of each of the forbidden 
  [\ion{O}{iii}]$\lambda\lambda$4959,\,5007 lines 
  is synthesized from a narrow core, as well as from a broad and blueshifted base 
  (e.g. Heckman et al. 1981,V\'{e}ron-Cetty et al. 2001, Zamanov et al. 2002, Christopoulou et al. 1997). 
  The atomic physical relationships 
  $F_{5007}/F_{4959} \doteq 3$(Storey \& Zeippen 2001) and $\lambda_{4959}/\lambda_{5007} = 0.9904$
  are employed to decrease the number of free parameters in the modelling of both narrow and broad components.
  2)As a second step the H$\beta$ line profile is synthesized from both a narrow Gaussian component and 
  a broader one. Although this representation can fit the H$\beta$ line core well and agree with a generally 
  accepted unified model for AGN, it could not fit the far blue wing of H$\beta$ adequately. Therefore, an 
  additional blueshifted Gaussian component with $\rm{FWHM \sim 6000\ km\ s^{-1}}$ should be
  acquired to fit the far blue wing of H$\beta$ (e.g. Sulentic et al. 2000c; Marziani et al. 2003b; 
  Korista \& Goad 2004). 
  In summary, a set of the following three Gaussian components are adopted to model the H$\beta$ profile: 
  a narrow core, a classical broad 
  component with FWHM$\sim1200\ \rm{km\ s^{-1}}$, and a very broad and blueshifted base. 
  The FWHM of narrow H$\beta$ is taken to be 
  equal to that of the [\ion{O}{iii}] core, because both components are emitted from the same region, i.e. from
  the narrow line region(NLR). The \ion{He}{ii}$\lambda$4684 line profile is easily fitted by a set of  
  two Gaussian profiles because of its substantial reflection in profile.
  As an illustration, the profile modelling of the spectrum observed at JD = 2,450,250.7 is shown schematically in Fig. 3.
  The observed profile is represented by a thin line, and the modelled profile by a solid line. 
  Each Gaussian component is shown by either a long- or short-dashed line.
  The residuals between the observed and modelled profile are presented in the bottom panel underneath the spectrum.

  \section{Results}

  The modelled flux of each component is calibrated by a constant total flux of [\ion{O}{iii}]$\lambda$5007,
  where $F([\ion{O}{iii}]\lambda5007) = (3.91 \pm 0.12) \times 10^{-13}\ \rm{ergs\ s^{-1} \ cm^{-2}}$ (Peterson
  et al. 2000). The small systematical flux difference between the two sets(set A and B) 
  are corrected by Formulas 5 and 6 in Peterson et al. (2000).

  The final results of profile modelling, along with the continuum and total H$\beta$ fluxes adopted from 
  Peterson et al. (2000), are given in Table 1. Column (1) lists the file name, and Column (2) the 
  corresponding Julian Date of observation. The continuum and H$\beta$ fluxes, both measured
  by Peterson et al. (2000), are listed in Columns (4) and (5), respectively. Column (6) is the flux of the subtracted 
  optical \ion{Fe}{ii} complex between rest wavelength 4434\AA \ and 4684\AA\ , along with the determined
  upper and lower limits. The modelled total flux of H$\beta$ is given in Column (7), and the total flux 
  of \ion{He}{ii}$\lambda$4686 in column (8). All the errors given in Columns (7) and (8) are caused
  by profile modelling.

  The correlation between the modelled flux of H$\beta$ 
  and the flux provided by Peterson et al. (2000) is illustrated in Fig. 4. The modelled H$\beta$ flux 
  containing all three components is represented by solid square symbols (Correlation I, for short), 
  and the modelled H$\beta$ flux in which the very broad component are excluded, by open triangles (Correlation 
  II, for short). Statistical analysis yields 
  a Spearman rank-order correlation coefficient $r_{s}=0.896$ ($P<10^{-4}$, where P is the probability that there is 
  null relation between two variables) for Correlation I and $r_{s}=0.708$($P<10^{-4}$) for Correlation II.
  Because of the lower inferred correlation coefficient for Correlation II with respect to that for Correlation I, 
  we clarify that the very broad component should not be ignored in accounting for the total flux of H$\beta$.
  Systematically, the modelled flux is slightly larger in this way than the flux provided by Peterson et
  al. (2000). This systematical enhancement can be explained easily by emission at the high velocity wing of the very broad base.
  In the study by Peterson et al. (2000), the H$\beta$ high velocity wing is truncated by integration ranging 
  from 4820\AA\ to 4910\AA\ in the observed frame. 
  Comparing the flux modelled independently in this paper to the one obtained in Peterson et al. (2000), we find a highly 
  significant, nearly linear correlation between them.

  \subsection{Light curves}

   The derived light curves of H$\beta$, \ion{He}{ii}$\lambda$4686, and \ion{Fe}{ii} are displayed in the bottom three
   panels in Fig. 5. Moreover, the light curves of H$\beta$ and continuum both derived by Peterson et al. (2000)
   are shown in the top two panels. The two independent H$\beta$ light curves are quite similar.
   The second panel, from bottom to top, shows the light curve of the \ion{He}{ii} line. Here, it should be emphasized that
   the flux of \ion{He}{ii} is obtained from the \ion{Fe}{ii} contamination-removed spectrum. 
   Because of the significant blending between \ion{He}{ii} and the \ion{Fe}{ii} complex, the \ion{Fe}{ii} emission is 
   a complicating factor when measuring the strength of \ion{He}{ii} line. In Fig. 5, the error bars overlaid
   on the \ion{He}{ii} light curve include only those uncertainties caused by profile modelling and do not reflect 
   the errors caused by the \ion{Fe}{ii} subtraction. The \ion{Fe}{ii} variations are shown in the bottom panel of Fig. 5.
   The length of each overlaid solid line corresponds to the range between upper and lower limits, which are
   determined by the iterative experiments of the \ion{Fe}{ii} subtraction. The pattern of the \ion{Fe}{ii} variations 
   closely follows the continuum (and H$\beta$) light curve. Despite all our efforts, 
   any significant lag of the \ion{Fe}{ii} complex with respect to the continuum could not be determined
   because of the large uncertainties of the \ion{Fe}{ii} flux.

   In the lower panel of Fig. 6, the flux of the \ion{Fe}{ii} complex 
   is plotted as a function of continuum flux as an additional test of line variability.
   The diagram shows a positive correlation between \ion{Fe}{ii} intensity and 
   continuum flux ($\rm{r_{s}=0.701,P<10^{-4}}$). The relationship of the H$\beta$ line 
   with respect to continuum is displayed in the upper panel ($\rm{r_{s}=0.564,P<10^{-4}}$).

   \subsection{Equivalent width and RFe}

   The Baldwin Effect (BE) defined as an anti-correlation between the equivalent width (EW) of \ion{C}{iv}
   $\lambda$1549 and continuum luminosity at $\lambda$1450 was first reported by Baldwin(1977).
   Subsequent observational studies indicated that the BE can be detected in almost all measurable high ionization 
   emission lines (Espey et al. 1993, Lanzetta et al. 1993, Zheng \& Malkan 1993) except \ion{N}{v}$\lambda$1240
   (Hamann \& Ferland 1999). Gilbert \& Peterson (2003) recently found a convincing intrinsic BE in the 
   broad H$\beta$ line of the active galaxy NGC\,5548 by analyzing spectra from International AGN 
   Watch collaboration (see also in Goad, Korista \& Knigge 2004). 

   In Fig. 7, the EW of the \ion{Fe}{ii} complex is plotted against continuum flux. There is a moderate correlation 
   between these data ($\rm{r_{s}=0.384},P<10^{-4}$), which confirms previous claims that no convincing BE 
   has been detected in the optical \ion{Fe}{ii} blends (Yee \& Oke 1981, Elston et al. 1994).  
   The relationship is influenced slightly by a few points with excessive \ion{Fe}{ii} contributions. These points clearly 
   deviate from the other points. Detailed inspection of the light curves indicates that 
   these specific points are mainly deduced from spectra observed from JD = 2,450,193.6 to 2,450,465.0.
   Subsequently, the corresponding original spectra, \ion{Fe}{ii} subtraction, and profile modelling
   are carefully inspected. We find that the deduced excessive \ion{Fe}{ii} emission is perhaps
   caused by the lower S/N ratio at blueward of the spectra. The bad quality of the spectra makes measurement of 
   the \ion{Fe}{ii} blends difficult and ultimately leads to an over-removal of the \ion{Fe}{ii} complex
   because of blending between \ion{Fe}{ii} and other faint emission lines, such as \ion{He}{i}$\lambda$4471.

   The intrinsic variations of $\rm{R_{Fe}}$ as a function of continuum flux are plotted in Fig. 8.
   $\rm{R_{Fe}}$ is one of the most important quantities describing the E1 parameter space, and it is defined as the flux ratio of 
   optical \ion{Fe}{ii} complex to H$\beta$. Although the flux of H$\beta$ includes the contributions of narrow, 
   broad, and very broad components, it should be noted that the flux of the narrow component of H$\beta$ is, of course, 
   expected to be constant. This plot shows a positive correlation between $\rm{R_{Fe}}$ and continuum flux logarithm. 
   The correlation coefficient derived by Spearman analysis is $\rm{r_{s}=0.521 (P<10^{-4})}$.
   An unweighted linear fit to this relation gives 
   $\rm{R_{Fe}}\propto (2.5\pm0.4)\log ([F_{\lambda}(5100\AA)/10^{-13} ergs\ s^{-1}\ cm^{-2}])$
   and is overplotted in Fig. 8 as a solid line. By analyzing the reverberation mapping (e.g. Blandford 
   \& Mckee 1982, Peterson et al.1998b) results of spectrophotometrical monitoring of a well-defined sample 
   of 17 Palomar Green quasars and 17 Seyfert galaxies, Kaspi et al. (2000) found that the relation between the Eddington ratio and 
   the continuum luminosity at $\lambda$5100 can be expressed as  $L/L_{\rm{E}} \propto [\lambda L_{\lambda}(5100\AA)/10^{44}
   \rm{ergs\ s^{-1}}]^{0.5}$. Consequently, by combining the above relationships, we find that $\rm{R_{Fe}}$ can be inferred 
   to scale with the Eddington ratio as $\rm{R_{Fe}} \propto (5.0\pm0.8) \log(L/M)$. Our relationship does not agree with that 
   found by Marziani et al. (2001, and references therein). To predict a grid of theoretical values in E1 parameter space,
   these authors obtained a global semi-theoretical relation between $\rm{R_{Fe}}$ and the Eddington ratio:
   $\rm{R_{Fe}}\propto 0.55\log(L/M)$. The discrepancy between the two relations is highly significant.

   \section{Discussion: Comparison of variability of \ion{Fe}{ii} complex: NGC\,4051 and other AGNs}

   In this paper, we detect a positive correlation between $\rm{R_{Fe}}$ and the continuum flux in NGC\,4051.
   However, it should be emphasized that the real situation is hard to handle when taking other results about 
   variations of \ion{Fe}{ii} emission into account. 
   For instance, by calculating the flux ratio of \ion{Fe}{ii} to H$\beta$,
   we find that in NGC\,7603, $\rm{R_{Fe}}$ apparently decreases with increasing continuum (Kollatschny 
   et al. 2000). This relation is displayed in Fig. 9. The point observed 
   at JD=24,044,168 is omitted because it clearly deviates from the other points. The difference in variability 
   behavior of $\rm{R_{Fe}}$ implies that the Seyfert galaxies NGC\,4051 and NGC\,7603 differ in those physical 
   conditions that govern the variability behavior of \ion{Fe}{ii} emission.

    In order to statistically investigate the behavior of \ion{Fe}{ii} variations, we collected some 
    results about \ion{Fe}{ii} variations from earlier publications. The comparison is summarized 
    in Table 2. Column 1 lists the  object name, and Column 2 the averaged FWHM of H$\beta$. Column 3 summarizes
    the relation between $\rm{R_{Fe}}$ and continuum flux. In fact, whether a positive 
    or a negative correlation can be obtained is determined by the fact that the variations of the \ion{Fe}{ii} blends
    are stronger or weaker in comparison to the H$\beta$ line. In Mark\,359, Mark\,1044, and Akn\,564, 
    the percentage variations
    in line fluxes of H$\beta$ and \ion{Fe}{ii}$\lambda$4550 were given by Giannuzzo \& Stirpe (1996). The variability
    behavior of $\rm{R_{Fe}}$ could be easily obtained in terms of the ratio between the percentage variation of \ion{Fe}{ii} 
    and that of H$\beta$. If the ratio is larger than unity, then $\rm{R_{Fe}}$ is expected 
    to increase with continuum. Insteady, $\rm{R_{Fe}}$ decreases with continuum when the ratio is less than 1.
    In NLS1 galaxy Mark\,110, $\rm{R_{Fe}}$ is expected to decrease with continuum flux both because the H$\beta$ line varied
    by a factor of about 2 and because the \ion{Fe}{ii} lines remained constant over the 10 years.

    By comparing the variability behaviors of different objects, we find that all objects with positive 
    correlations have narrow H$\beta$
    profiles and can be classified as NLS1s. In contrast, the remaining two sources with negative correlations 
    have relatively broad H$\beta$ profiles (i.e. $\rm{FWHM>1500\ km\ s^{-1}}$). The dichotomy 
    in variability behavior of $\rm{R_{Fe}}$ suggests  
    that the variability amplitude of the \ion{Fe}{ii} complex in Seyfert galaxies might be
    correlated with the width of the H$\beta$ line. Because the standard photoionization models cannot 
    interpret the strong \ion{Fe}{ii} emission, we attempt to interpret the dichotomy in variability
    behavior of $\rm{R_{Fe}}$ in the framework of collisional models, in which the bulk excitation of the
    optical \ion{Fe}{ii} lines is due to collisional excitation in a high density optically thick cloud
    illuminated and heated mainly by X-rays photons (Wills, Netzer \& Wills 1985, Kwan et al.1995, Sigut \& Pradhan 
    2003, Verner et al. 1999, Collin-Souffrin et al. 1986, 1988).
    The \ion{Fe}{ii} emission region is typical of $N_{e}\sim10^{10-12}\rm{cm^{-3}}$,
    $N_{\rm{H}}>10^{24}\rm{cm^{-2}}$, and $T_{e}\sim8000$K.

    We discuss the observed \ion{Fe}{ii} variability in terms of the extensively used line responsivity     
    $\partial j_{l}(t)/\partial F_{c}$, where $j_{l}$ is the emissivity of a given line and $F_{c}$  
    the incident ionizing continuum flux. The time dependent responsivity means that the gas requires 
    some time to equilibrate to a new continuum level. We clarify that this delay, in minutes, 
    is so short that it can be neglected on the basis of the following discussions. 
    Relaxation to thermal balance takes place on the timescale 
    $t_{\rm{cool}}\sim5\times10^{11}n^{-1}_{e}\ \rm{s}$ (Krolik 1999), where the fact that the cooling function
    is usually $\sim10^{-23}\ \rm{erg\ cm^{3}\ s^{-1}}$ when the temperature is around $10^{4}$K is used.
    This formula provides a cooling timescale of about 0.1-1000 seconds when typical values of density
    in BLR are taken (e.g. in NGC\,4051 $N_{e}\sim10^{10}\ \rm{cm^{-3}}$ (Hyung et al. 2000), $8.7<\log N_e<9.1$
    (Komossa \& Mathur 2001)). These calculations indicate that, relative to the timescales on which 
    the intrinsic continuum changes (weeks to years), the cooling timescale can be entirely ignored.
    
    Now we focus attention on the 
    line responsivity $\partial j_{l}/\partial F_{c}$, and consider the two possible explanations:
    
    \begin{enumerate}

    \item \emph{Line responsivity as a function of electron density.} Recent numerical calculations have indicated 
    that the responsivity of the \ion{Fe}{ii} line flux in higher density case is much larger than in 
    the case with lower density (see Fig.6 in Sigut, Pradhan \& Nahar (2004)). The
    \ion{Fe}{ii} flux is enhanced by about one order in the model with $\log n_{\rm{H}}=9.5$, but by
    nearly two orders in the model with $\log n_{\rm{H}}=11.5$, when the ionization parameter increases
    from $10^{-3}$ to $10^{-1.5}$. Since there is a significant correlation between $\rm{R_{Fe}}$ and 
    electron density (Aoki \& Yoshida 1999; Wills et al. 1999; Marziani et al. 2001 and references therein), adopting 
    the generally accepted E1 correlation $\rm{R_{Fe}}\propto\rm{FWHM}^{-1}$ yields $\rm{FWHM(H\beta)\propto n^{-1}_e}$.
    For example, in I\,ZW1 the line ratio  \ion{Si}{iii}]$\lambda$1892/\ion{C}{iii}]$\lambda$1909
    $\approx$3.5 (Laor et al. 1997) is much larger than the typical value for quasars ($\approx$0.3, 
    Laor et al. 1995). The ratio \ion{Si}{iii}]/\ion{C}{iii}] is a useful density diagnostic in BLR (Ferland et al. 2000).
    In general, the responsivity of the \ion{Fe}{ii} lines in NLS1 is consequently expected to be larger than 
    in Broad-Line Seyfert 1 galaxy(BLSy1).
   
    \item \emph{Line responsivity as a function of incident continuum shape.} It is clear that the 
    reprocessed spectrum also depends on the shape of incident continuum. Continuum energies that
    should most affect \ion{Fe}{ii} strength are $h\nu>$800eV(Krolik \& Kallman 1988). 
    Recently, the positive
    correlations between ROSAT HR1 and Count Rates were identified for six out of eight NLS1s,
    but the anti-correlations were identified for seven out of 14 BLSy1s(Cheng et al. 2002). There
    were no detectable correlations in the other two NLS1s and 7 BLSy1s. This means that, in general, 
    the fraction of ionizing 
    photons at high energy level increases with incident continuum in NLS1. In contrast, a
    decreased fraction can usually be found in BLSy1. If so, the magnitude of changes in heat contributed 
    by the high energy photons are stronger in NLS1 and weaker in BLSy1, so that, the observed variations
    of the optical \ion{Fe}{ii} blends necesserialy become strong in NLS1 and weak in BLSy1.  
    \end{enumerate}

    In summary, the dichotomy in the variability behavior of $\rm{R_{Fe}}$ could be caused by one of two
    different physical conditions, i.e. either by electron density in a single cloud or by variability behavior of 
    incident high energy photons. At present, however, evidence is not conclusive enough to
    determine which of them is more important. Although we discuss them separately, it is also possible 
    for both mechanisms to act together in AGN. It might be worthwhile to extend variability campaigns to larger Seyfert 
    samples with different broad line widths to investigate the validity of the trend mentioned above.  
    It is also likely that complicated photoionization model calculations 
    are necessary to distinguish between the proposed two interpretations. However, this model study is 
    beyond the scope of this paper.

%

\section{Conclusions}

    We perform new analysis of the archival spectra of NGC\,4051 extracted from the AGN Watch project
    in order to investigate the variability of optical \ion{Fe}{ii} emission. The template of BG92 is 
    used to remove and to measure the \ion{Fe}{ii} complex. The other emission lines are profiled by 
    multi-component profile modelling. This analysis allows us to make the following conclusions:

   \begin{enumerate}
   
   \item In NGC\,4051, we find that the optical \ion{Fe}{ii} complex was variable during the three years period.
         The \ion{Fe}{ii} variations closely follow the continuum variations, and the intensity of \ion{Fe}{ii}
         evidently increases with the continuum flux.
   
   \item A positive correlation between the EW of \ion{Fe}{ii} and the continuum flux is
         identified in NGC\,4051. This result agrees with the previous claims that
         no convincing BE of \ion{Fe}{ii} has been detected untill now.

   \item By comparing the variations of H$\beta$ and \ion{Fe}{ii}, a positive correlation between 
         $\rm{R}_{Fe}$ and continuum flux is obtained in NGC\.4051. The unweighted
         fit gives the relation $\rm{R_{Fe}}\propto(5.0\pm0.8)\log(L/M)$. 
         This relation is significantly different from the relation $\rm{R_{Fe}}\propto0.55\log(L/M)$, 
         predicted by the semi-theoretical formula (Marziani et al. 2000).

   \item We find an inverse correlation between $\rm{R}_{Fe}$ and continuum in Seyfert galaxy 
         NGC\,7603 (Kollatschny et al. 2000). The difference in the variability behavior of $\rm{R}_{Fe}$
         implies that the Seyfert galaxies NGC\,4051 and NGC\,7603 differ in physical conditions 
         governing the variability of the optical \ion{Fe}{ii} blends.
         Furthermore, in six selected Seyfert
         galaxies, the positive correlations are identified in 4 out of 5 NLS1s and for the negative 
         correlations, in the remaining two objects whose H$\beta$ profiles are relatively broad.
         The different electron density of BLR clouds and variability behavior of high energy photons
         are put forward to interpret the dichotomy in variability behavior of $\rm{R_{Fe}}$.

   \end{enumerate}

\begin{acknowledgements}
      We are grateful to the anonymous referee for many useful suggestions.
      The authors acknowledge many valuable discussions with Dr. Xu, D. W., Hao, C. N., S. Komossa, and
      Mao, Y. F. This work has made use of the archival spectroscopic data of the AGN Watch campaign.
      We thank Prof. Bradley M. Peterson for providing a grant for using the spectra of NGC\,4051.
      Our thanks also go to Dr. Todd A. Boroson and Richard F. Green for providing the \ion{Fe}{ii} template.
      This work was financially supported by the Ministry of Science and Technology of China, 
      under grant NKBRSF G19990754, and by the NSF of China, No. 10473013.

\end{acknowledgements}

\clearpage
 
  \begin{figure*}
  \includegraphics[width=13cm]{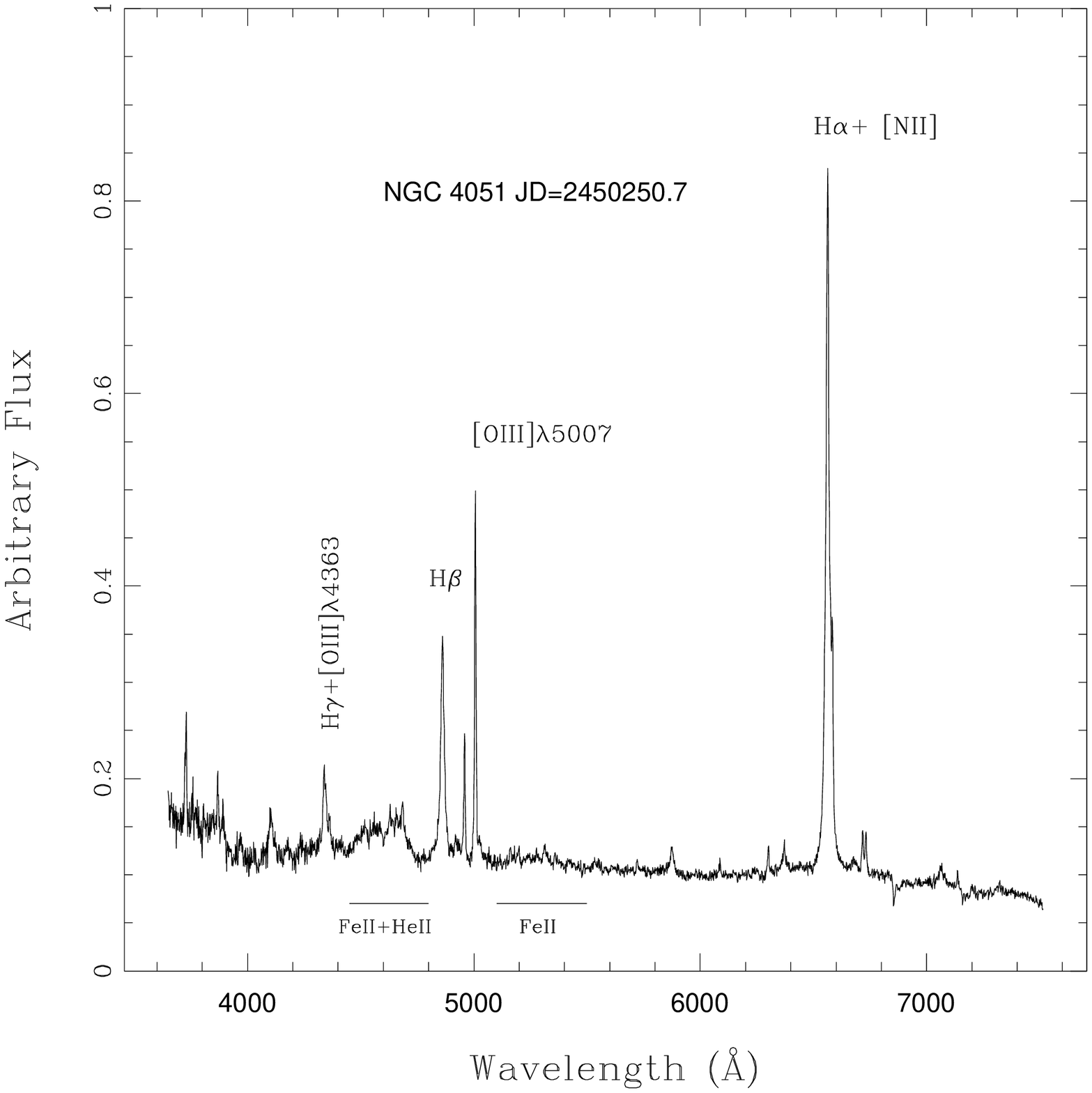}
  \caption{--- Rest frame and Galactic extinction corrected spectrum of NGC\,4051 taken on
  June 16, 1996 (JD = 2,450,250.7). The most prominent
  emission lines and optical \ion{Fe}{ii} complex are marked.}
  \end{figure*}

  \begin{figure*}
  \includegraphics[width=13cm]{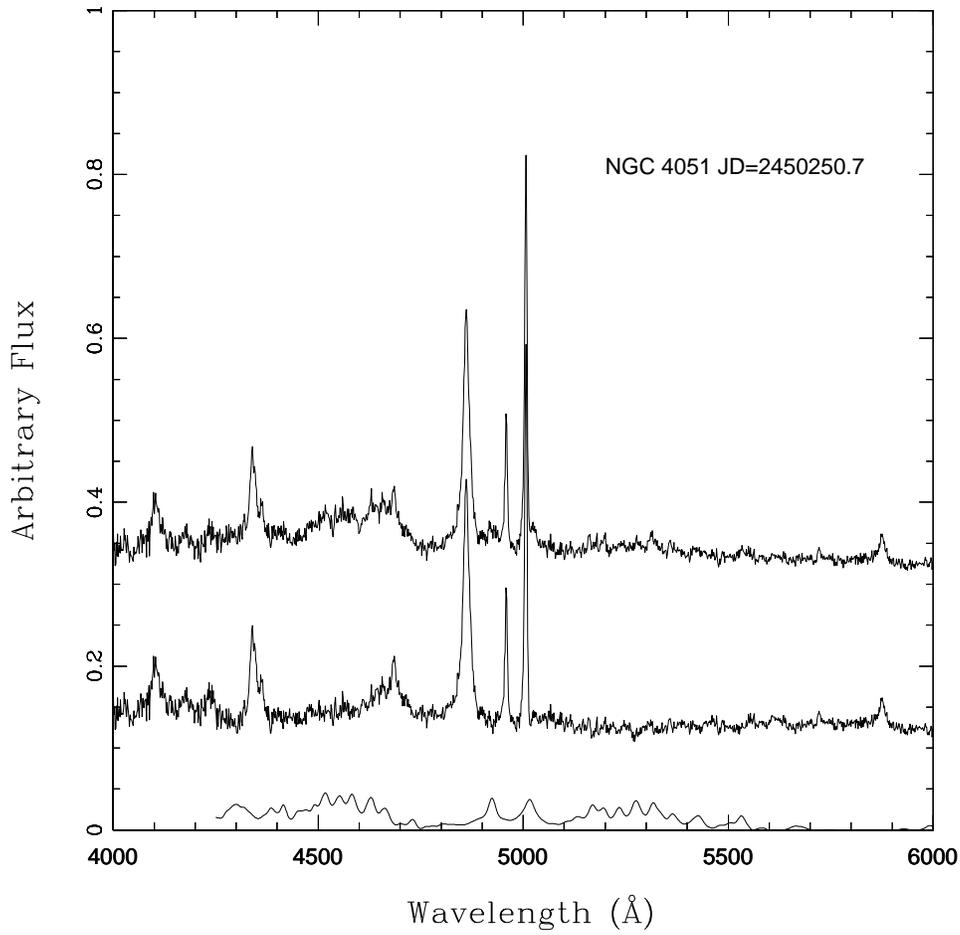}
  \caption{--- The scheme of the \ion{Fe}{ii} complex subtraction of spectrum taken at JD = 2,450,250.7.
   The top and middle curves are \ion{Fe}{ii} emission blended and subtracted spectra, respectively.
   The observed spectrum is shifted upward by an arbitrary amount. The bottom spectrum
   is the best adopted template of the  \ion{Fe}{ii} complex.}
  \end{figure*}

  \begin{figure*}
  \includegraphics[width=13cm]{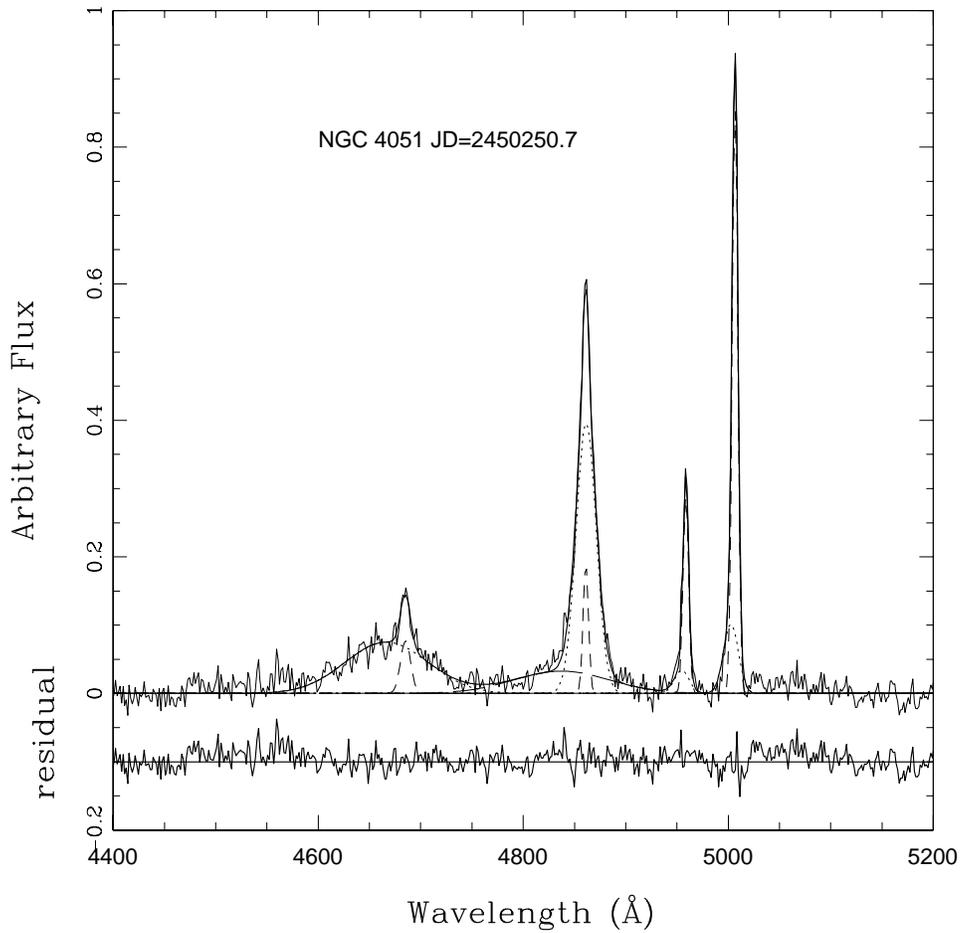}
  \caption{--- An illustration of line-profile modelling of a continuum-removed spectrum 
   taken on June 16, 1996 (JD = 2,450,250.7). The observed profile
   is shown by the thin solid line, and the modelled profile, by the thick solid line. The narrow and
   broad components of each emission feature are represented by long and short dashed lines, respectively.  
   The residuals of profile modelling are displayed in the lower panel.
   }
  \end{figure*}

  \begin{figure}
  \includegraphics[width=13cm]{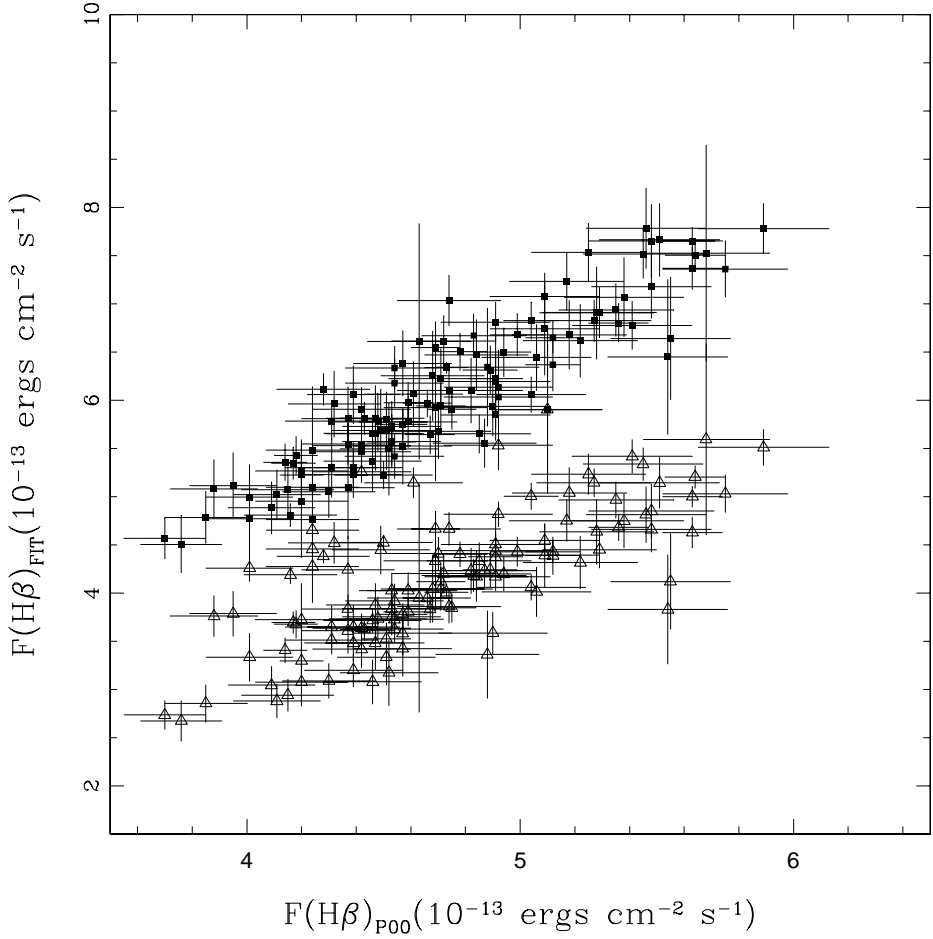}
  \caption{--- The correlation between two independent H$\beta$ measurements(see text for details).
   The horizontal axis is the
   value obtained by Peterson et al. (2000), and the vertical one the value provided by profile modelling.
   The total flux including
   all three components is denoted by solid squares ($r_{s}=0.896, P<10^{-4}$), while the flux, including a narrow peak
   and a classical broad component, is denoted by open triangles ($r_{s}=0.708, P<10^{-4}$).}
  \end{figure}

   \begin{figure*}
   \centering
   \sidecaption
   \includegraphics[width=16cm]{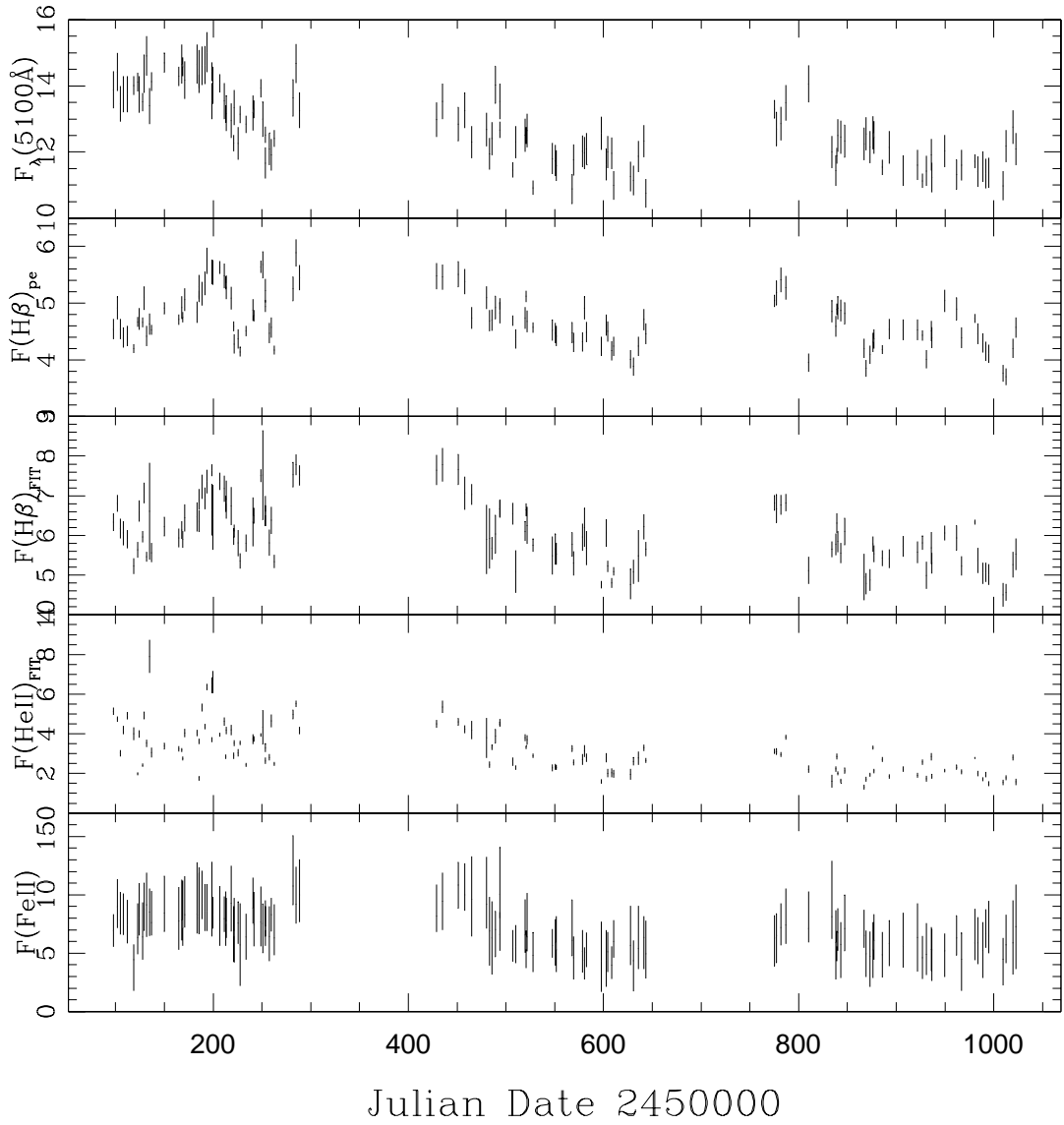}
   \caption{--- Light curves of continuum and emission lines for NGC\,4051 with 120 points taken between Jan 15, 1996 and
   Jul 28, 1998. The first panel shows the light curve of continuum in units of $10^{-15}\ \rm{ergs\ s^{-1}\ cm^{-2}\ \AA}$.
   The H$\beta$ light curve adopted from Peterson et al. (2000) is displayed in the second panel in units of 
   $10^{-13}\ \rm{ergs\ s^{-1}\ cm^{-2}}$. The third, fourth, and fifth 
   panels illustrate, respectively, the emission line light curves of H$\beta$, \ion{He}{ii}$\lambda$4686, and \ion{Fe}{ii} in units of
   $10^{-13}\ \rm{ergs\ s^{-1}\ cm^{-2}}$.}
   \end{figure*}

  \begin{figure}
   \centering
   \sidecaption
   \includegraphics[width=13cm]{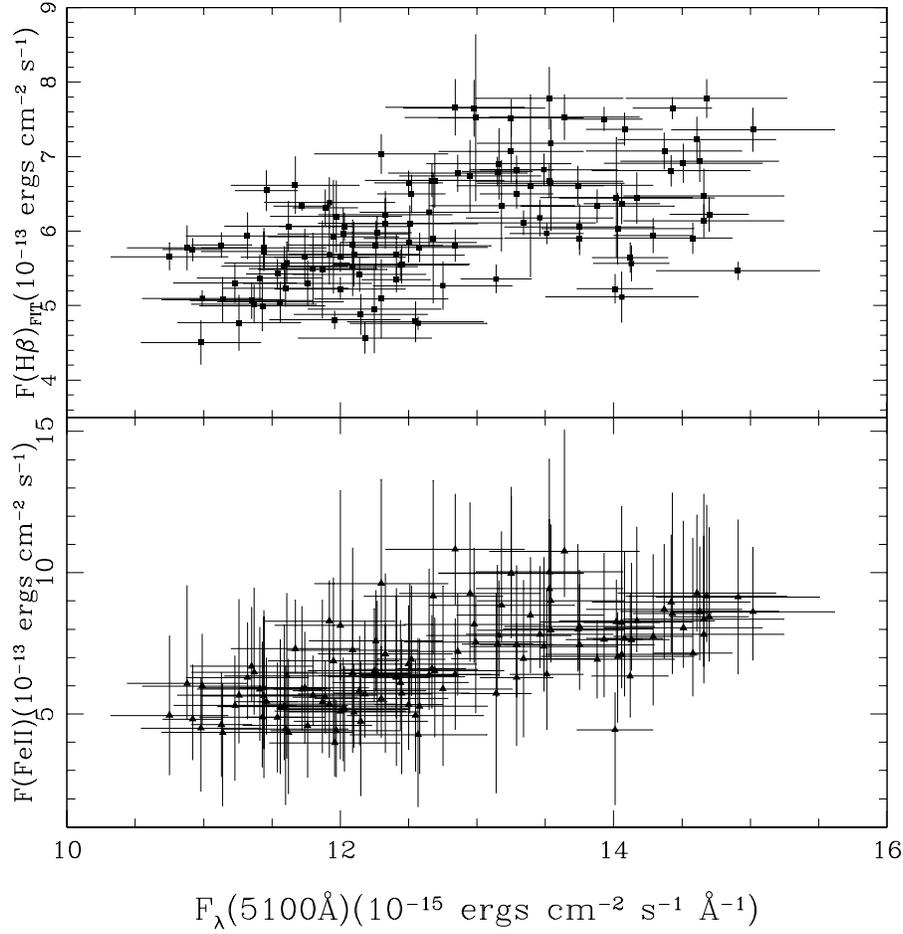}
   \caption{--- Lower panel: line intensity of the optical \ion{Fe}{ii} blends plotted against
    continuum at rest wavelength 5100\AA\ ($r_{s}=0.701, P<10^{-4}$). In addition, intensity of H$\beta$ 
    as function of continuum flux is shown in the upper panel ($r_{s}=0.564, P<10^{-4}$).} 
   \end{figure}

  \begin{figure}
   \centering
   \sidecaption
   \includegraphics[width=13cm]{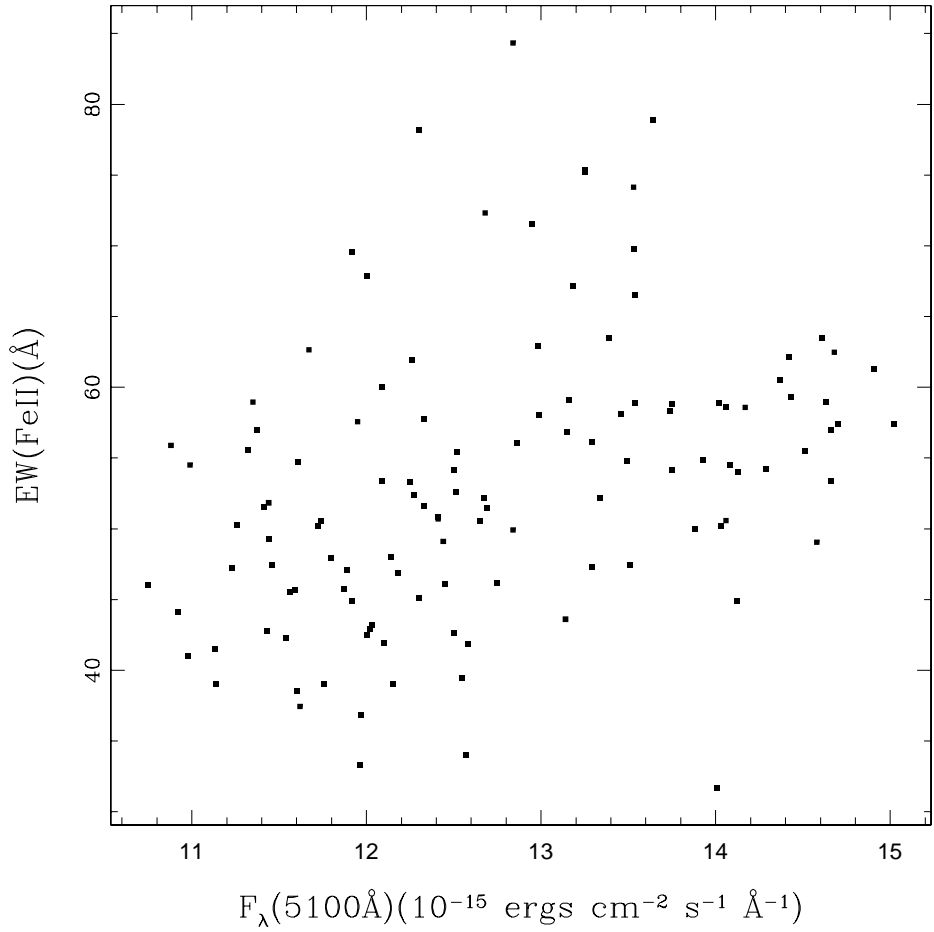}
   \caption{--- A plot of equivalent width for the optical \ion{Fe}{ii} complex against the flux of continuum
   ($r_{s}=0.384, P<10^{-4}$).}
   \end{figure}

    \begin{figure}
   \centering
   \sidecaption
   \includegraphics[width=13cm]{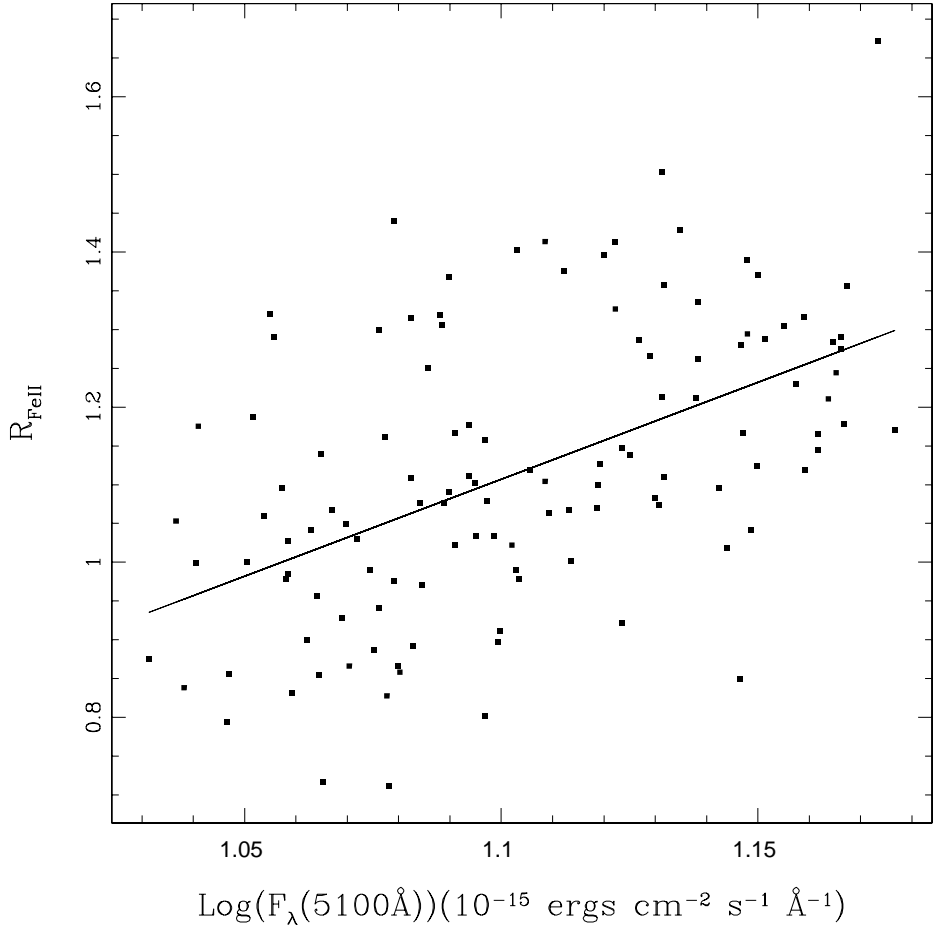}
   \caption{--- $\rm{R_{Fe}}$ plotted against the logarithm of continuum flux ($r_{s}=0.521, P<10^{-4}$). 
   The best fitted relation $\rm{R_{Fe}}\propto(5.0\pm0.8)\log(L/M)$ is shown by the overlaid solid curve.} 
   \end{figure}

   \begin{figure}
   \centering
   \sidecaption
   \includegraphics[width=13cm]{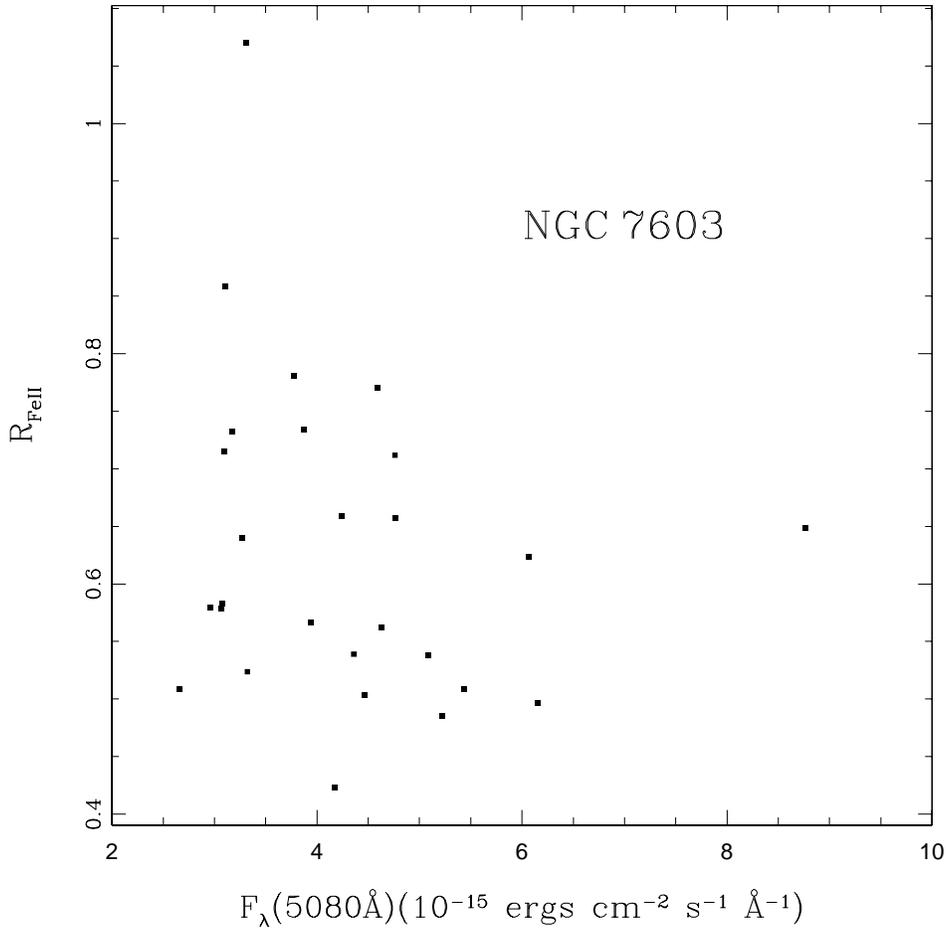}
   \caption{--- A plot of $\rm{R_{Fe}(4570\AA)}$ against the continuum flux in NGC\,7603. The fluxes are
   measured by Kollatschny et al.(2000). $\rm{R_{Fe}}$ is calculated by flux ratio between
   \ion{Fe}{ii} and H$\beta$.}
   \end{figure}

\clearpage

   \begin{longtable}{cccccccc}
   \caption{Continuum and integrated line fluxes.}\\
   \hline
   \hline
   File name & $\rm{Julian\ Date}^{a}$ & Set & $F_\lambda(5100\AA)^{\rm{b}}$ & $F\rm{(H\beta_{P})}^{\mathrm{c}}$
   & $F(\ion{Fe}{ii})^{\rm{c}}$ & $F\rm{(H\beta)}^{\rm{d,e}}$ & $F\rm{(\ion{He}{ii}\lambda4686)}^{\mathrm{d,f}}$\\
   (1)&(2)&(3)&(4)&(5)&(6)&(7)&(8)\\
   \hline
   \endfirsthead
   \caption{continued.}\\
   \hline
   \hline
   File name & $\rm{Julian\ Date}^{\rm{a}}$ & Set & $F_\lambda(5100\AA)^{\rm{b}}$ & $F\rm{(H\beta_{P})}^{\mathrm{c}}$
   & $F(\ion{Fe}{ii})^{\rm{c}}$ & $F\rm{(H\beta)}^{\rm{d,e}}$ & $F\rm{(\ion{He}{ii}\lambda4686)}^{\mathrm{d,f}}$\\
   (1)&(2)&(3)&(4)&(5)&(6)&(7)&(8)\\

   \hline
   \endhead
   \hline
   \endfoot
   n00098b & 98.0 \dotfill & B & $13.88\pm0.56$ & $4.54\pm0.18$ & $6.94^{+1.39}_{-1.39}$ & $6.34\pm0.23$ & $5.13\pm0.19$\\
   n00102b & 102.0\dotfill & B & $14.42\pm0.58$ & $4.91\pm0.20$ & $8.96^{+2.39}_{-1.79}$ & $6.81\pm0.21$ & $4.74\pm0.15$\\
   n00105b & 105.1\dotfill & B & $13.46\pm0.54$ & $4.54\pm0.18$ & $7.82^{+2.41}_{-1.20}$ & $6.18\pm0.25$ & $3.02\pm0.16$\\
   n00108b & 108.0\dotfill & B & $13.75\pm0.55$ & $4.39\pm0.17$ & $8.09^{+2.02}_{-1.52}$ & $6.06\pm0.30$ & $4.20\pm0.21$\\
   n00112b & 112.0\dotfill & B & $13.75\pm0.55$ & $4.42\pm0.18$ & $7.45^{+2.13}_{-1.60}$ & $5.91\pm0.23$ & $4.92\pm0.18$\\
   n00118a & 118.8\dotfill & A & $14.01\pm0.28$ & $4.20\pm0.08$ & $4.44^{+1.33}_{-2.66}$ & $5.23\pm0.20$ & $4.01\pm0.32$\\
   n00122a & 122.8\dotfill & A & $14.12\pm0.28$ & $4.67\pm0.09$ & $6.34^{+2.93}_{-1.45}$ & $5.64\pm0.20$ & $2.00\pm0.08$\\
   n00123b & 124.0\dotfill & B & $13.74\pm0.55$ & $4.72\pm0.19$ & $8.01^{+3.00}_{-1.50}$ & $6.61\pm0.27$ & $4.00\pm0.19$\\
   n00127a & 127.8\dotfill & A & $13.51\pm0.27$ & $4.66\pm0.09$ & $6.41^{+2.96}_{-1.97}$ & $5.97\pm0.14$ & $2.42\pm0.09$\\
   n00129b & 129.0\dotfill & B & $14.37\pm0.57$ & $5.09\pm0.20$ & $8.70^{+2.32}_{-1.74}$ & $7.07\pm0.25$ & $4.93\pm0.18$\\
   n00132b & 132.0\dotfill & B & $14.91\pm0.60$ & $4.42\pm0.18$ & $9.14^{+2.74}_{-2.74}$ & $5.47\pm0.12$ & $3.52\pm0.20$\\
   n00134b & 134.9\dotfill & B & $13.39\pm0.54$ & $4.63\pm0.19$ & $8.50^{+2.04}_{-2.04}$ & $6.61\pm1.22$ & $7.90\pm0.83$\\
   n00136a & 136.8\dotfill & A & $14.13\pm0.28$ & $4.53\pm0.09$ & $7.63^{+2.72}_{-1.09}$ & $5.57\pm0.24$ & $3.07\pm0.25$\\
   n00149a & 149.8\dotfill & A & $14.70\pm0.29$ & $4.91\pm0.10$ & $8.44^{+3.17}_{-1.59}$ & $6.23\pm0.23$ & $3.38\pm0.14$\\
   n00164a & 164.8\dotfill & A & $14.29\pm0.29$ & $4.71\pm0.09$ & $7.75^{+2.91}_{-2.45}$ & $5.94\pm0.24$ & $3.24\pm0.13$\\
   n00167b & 167.7\dotfill & B & $14.66\pm0.59$ & $4.92\pm0.20$ & $7.82^{+3.48}_{-1.74}$ & $6.14\pm0.22$ & $3.18\pm0.09$\\
   n00168a & 168.8\dotfill & A & $14.58\pm0.29$ & $4.75\pm0.09$ & $7.15^{+4.08}_{-1.53}$ & $5.91\pm0.21$ & $2.76\pm0.09$\\
   n00170b & 170.8\dotfill & B & $14.17\pm0.57$ & $5.06\pm0.20$ & $8.30^{+3.32}_{-1.11}$ & $6.44\pm0.35$ & $4.04\pm0.21$\\
   n00183b & 183.6\dotfill & B & $14.66\pm0.59$ & $4.84\pm0.19$ & $8.35^{+4.45}_{-1.67}$ & $6.47\pm0.37$ & $4.05\pm0.16$\\
   n00185b & 185.6\dotfill & B & $14.51\pm0.58$ & $5.29\pm0.21$ & $8.05^{+3.79}_{-1.42}$ & $6.91\pm0.26$ & $3.62\pm0.13$\\
   n00185a & 185.7\dotfill & A & $14.06\pm0.28$ & $5.12\pm0.10$ & $8.24^{+4.12}_{-1.55}$ & $6.37\pm0.27$ & $1.73\pm0.09$\\
   n00188b & 188.6\dotfill & B & $14.61\pm0.58$ & $5.17\pm0.21$ & $9.28^{+2.78}_{-1.36}$ & $7.23\pm0.30$ & $5.33\pm0.14$\\
   n00191b & 191.6\dotfill & B & $14.63\pm0.58$ & $5.35\pm0.21$ & $8.63^{+2.30}_{-1.73}$ & $6.94\pm0.28$ & $4.37\pm0.15$\\
   n00193b & 193.6\dotfill & B & $15.02\pm0.60$ & $5.75\pm0.23$ & $8.62^{+2.30}_{-1.73}$ & $7.36\pm0.29$ & $6.35\pm0.16$\\
   n00198a & 198.7\dotfill & A & $14.43\pm0.29$ & $5.63\pm0.11$ & $8.56^{+4.28}_{-2.14}$ & $7.65\pm0.15$ & $3.70\pm0.12$\\
   n00198b & 198.8\dotfill & B & $13.54\pm0.54$ & $5.55\pm0.22$ & $9.01^{+2.70}_{-1.80}$ & $6.64\pm0.64$ & $6.44\pm0.40$\\
   n00199b & 199.6\dotfill & B & $14.02\pm0.56$ & $5.54\pm0.22$ & $8.26^{+1.50}_{-1.13}$ & $6.45\pm0.80$ & $6.61\pm0.55$\\
   n00206a & 206.7\dotfill & A & $14.08\pm0.28$ & $5.63\pm0.11$ & $7.67^{+3.07}_{-2.05}$ & $7.37\pm0.22$ & $3.95\pm0.10$\\
   n00211b & 211.6\dotfill & B & $13.54\pm0.54$ & $5.48\pm0.22$ & $7.97^{+1.90}_{-1.14}$ & $7.18\pm0.33$ & $4.62\pm0.21$\\
   n00212a & 212.8\dotfill & A & $13.15\pm0.26$ & $5.36\pm0.11$ & $7.48^{+2.80}_{-1.87}$ & $6.80\pm0.20$ & $2.84\pm0.11$\\
   n00213b & 213.6\dotfill & B & $13.16\pm0.53$ & $5.28\pm0.21$ & $7.78^{+1.95}_{-1.46}$ & $6.91\pm0.48$ & $4.17\pm0.19$\\
   n00218b & 218.6\dotfill & B & $12.95\pm0.52$ & $5.09\pm0.20$ & $9.27^{+3.22}_{-2.42}$ & $6.74\pm0.48$ & $4.20\pm0.26$\\
   n00220a & 220.8\dotfill & A & $12.27\pm0.25$ & $4.59\pm0.09$ & $6.43^{+2.57}_{-2.14}$ & $5.98\pm0.21$ & $2.90\pm0.17$\\
   n00221b & 221.7\dotfill & B & $13.34\pm0.53$ & $4.28\pm0.17$ & $6.96^{+2.78}_{-2.78}$ & $6.11\pm0.17$ & $3.53\pm0.14$\\
   n00225b & 225.6\dotfill & B & $12.26\pm0.49$ & $4.37\pm0.17$ & $7.59^{+1.79}_{-1.79}$ & $5.81\pm0.33$ & $3.04\pm0.18$\\
   n00227a & 227.8\dotfill & A & $13.14\pm0.26$ & $4.14\pm0.08$ & $5.73^{+3.53}_{-3.53}$ & $5.36\pm0.19$ & $3.54\pm0.11$\\
   n00233a & 233.7\dotfill & A & $12.84\pm0.26$ & $4.51\pm0.09$ & $6.41^{+1.97}_{-1.97}$ & $5.81\pm0.22$ & $2.43\pm0.11$\\
   n00240b & 240.6\dotfill & B & $13.18\pm0.53$ & $4.88\pm0.19$ & $8.85^{+2.62}_{-1.31}$ & $6.34\pm0.61$ & $3.73\pm0.25$\\
   n00241a & 241.7\dotfill & A & $13.29\pm0.27$ & $4.78\pm0.10$ & $7.46^{+2.80}_{-1.87}$ & $6.50\pm0.20$ & $3.74\pm0.74$\\
   n00248a & 248.7\dotfill & A & $13.93\pm0.28$ & $5.64\pm0.11$ & $7.64^{+3.06}_{-2.04}$ & $7.50\pm0.16$ & $3.95\pm0.09$\\
   n00250b & 250.7\dotfill & B & $12.99\pm0.52$ & $5.68\pm0.23$ & $7.54^{+1.67}_{-2.51}$ & $7.52\pm1.12$ & $4.34\pm0.85$\\
   n00253b & 253.6\dotfill & B & $11.67\pm0.47$ & $5.22\pm0.21$ & $7.31^{+1.72}_{-0.86}$ & $6.62\pm0.38$ & $3.29\pm0.20$\\
   n00253a & 253.7\dotfill & A & $11.67\pm0.47$ & $5.22\pm0.21$ & $6.94^{+2.60}_{-1.30}$ & $6.50\pm0.26$ & $2.64\pm0.16$\\
   n00257b & 257.7\dotfill & B & $12.09\pm0.48$ & $4.47\pm0.18$ & $6.45^{+2.58}_{-2.15}$ & $5.82\pm0.33$ & $2.81\pm0.16$\\
   n00259b & 259.6\dotfill & B & $11.92\pm0.48$ & $4.57\pm0.18$ & $8.29^{+1.44}_{-1.44}$ & $6.38\pm0.34$ & $4.66\pm0.33$\\
   n00262a & 262.7\dotfill & A & $12.41\pm0.25$ & $4.17\pm0.08$ & $6.29^{+2.90}_{-1.45}$ & $5.35\pm0.17$ & $2.49\pm0.10$\\
   n00281b & 281.6\dotfill & B & $13.64\pm0.55$ & $5.25\pm0.21$ &$10.76^{+4.31}_{-1.62}$ & $7.53\pm0.31$ & $4.98\pm0.22$\\
   n00284b & 284.7\dotfill & B & $14.68\pm0.59$ & $5.89\pm0.24$ & $9.17^{+3.24}_{-1.62}$ & $7.78\pm0.26$ & $5.52\pm0.16$\\
   n00288b & 288.6\dotfill & B & $13.25\pm0.53$ & $5.45\pm0.22$ & $9.97^{+3.07}_{-2.30}$ & $7.52\pm0.26$ & $4.17\pm0.19$\\
   n00429b & 429.0\dotfill & B & $12.98\pm0.52$ & $5.48\pm0.23$ & $8.17^{+2.72}_{-1.36}$ & $7.65\pm0.38$ & $4.50\pm0.20$\\
   n00435b & 435.0\dotfill & B & $13.53\pm0.54$ & $5.46\pm0.22$ & $9.44^{+2.46}_{-2.46}$ & $7.78\pm0.42$ & $5.36\pm0.32$\\
   n00451b & 451.0\dotfill & B & $12.84\pm0.51$ & $5.51\pm0.22$ &$10.83^{+1.97}_{-1.97}$ & $7.66\pm0.38$ & $4.60\pm0.18$\\
   n00458b & 458.0\dotfill & B & $13.25\pm0.53$ & $5.38\pm0.22$ & $9.99^{+2.72}_{-1.36}$ & $7.07\pm0.41$ & $4.23\pm0.20$\\
   n00465b & 465.0\dotfill & B & $12.30\pm0.49$ & $4.74\pm0.19$ & $9.62^{+3.66}_{-3.20}$ & $7.04\pm0.27$ & $4.20\pm0.46$\\
   n00480b & 480.0\dotfill & B & $12.68\pm0.51$ & $5.10\pm0.20$ & $9.17^{+4.08}_{-2.04}$ & $5.90\pm0.87$ & $3.80\pm1.00$\\
   n00483b & 483.1\dotfill & B & $11.95\pm0.48$ & $4.69\pm0.19$ & $6.88^{+2.95}_{-2.95}$ & $5.92\pm0.76$ & $2.44\pm0.17$\\
   n00485b & 485.9\dotfill & B & $12.41\pm0.50$ & $4.70\pm0.19$ & $6.31^{+3.15}_{-3.15}$ & $5.68\pm0.28$ & $3.30\pm0.12$\\
   n00488b & 489.0\dotfill & B & $14.03\pm0.56$ & $4.92\pm0.20$ & $7.04^{+1.56}_{-2.34}$ & $6.03\pm0.48$ & $3.88\pm0.35$\\
   n00493a & 493.1\dotfill & A & $12.67\pm0.25$ & $4.99\pm0.10$ & $6.61^{+1.89}_{-1.42}$ & $6.68\pm0.22$ & $4.59\pm0.17$\\
   n00494b & 494.0\dotfill & B & $13.53\pm0.54$ & $4.83\pm0.19$ &$10.03^{+4.01}_{-2.01}$ & $6.67\pm0.23$ & $4.48\pm0.13$\\
   n00506a & 506.8\dotfill & A & $11.46\pm0.23$ & $4.69\pm0.09$ & $5.44^{+1.56}_{-1.17}$ & $6.55\pm0.27$ & $2.60\pm0.21$\\
   n00509b & 510.0\dotfill & B & $12.30\pm0.49$ & $4.37\pm0.17$ & $5.55^{+1.85}_{-1.39}$ & $5.09\pm0.53$ & $2.30\pm0.12$\\
   n00519b & 519.8\dotfill & B & $12.51\pm0.50$ & $4.74\pm0.19$ & $6.58^{+3.03}_{-1.56}$ & $6.10\pm0.24$ & $3.80\pm0.17$\\
   n00520a & 520.8\dotfill & A & $12.50\pm0.25$ & $5.12\pm0.10$ & $5.33^{+1.60}_{-1.60}$ & $6.65\pm0.16$ & $3.32\pm0.09$\\
   n00521b & 521.8\dotfill & B & $12.65\pm0.51$ & $4.68\pm0.19$ & $6.39^{+3.76}_{-1.24}$ & $6.25\pm0.47$ & $3.70\pm0.24$\\
   n00527a & 527.8\dotfill & A & $10.92\pm0.22$ & $4.57\pm0.09$ & $4.82^{+1.93}_{-1.45}$ & $5.75\pm0.16$ & $2.90\pm0.11$\\
   n00547b & 547.6\dotfill & B & $11.80\pm0.47$ & $4.52\pm0.18$ & $5.66^{+1.41}_{-1.06}$ & $5.49\pm0.48$ & $2.29\pm0.17$\\
   n00550b & 550.8\dotfill & B & $11.74\pm0.47$ & $4.47\pm0.18$ & $5.93^{+1.98}_{-1.98}$ & $5.65\pm0.38$ & $2.34\pm0.15$\\
   n00551b & 551.8\dotfill & B & $11.59\pm0.46$ & $4.42\pm0.18$ & $5.29^{+2.88}_{-1.92}$ & $5.53\pm0.27$ & $2.32\pm0.13$\\
   n00567b & 567.6\dotfill & B & $10.88\pm0.44$ & $4.48\pm0.18$ & $6.08^{+3.47}_{-1.30}$ & $5.77\pm0.30$ & $3.25\pm0.16$\\
   n00569b & 569.6\dotfill & B & $11.76\pm0.47$ & $4.31\pm0.17$ & $4.59^{+1.84}_{-1.84}$ & $5.30\pm0.31$ & $2.56\pm0.16$\\
   n00578b & 578.6\dotfill & B & $12.02\pm0.48$ & $4.32\pm0.17$ & $5.16^{+1.82}_{-1.82}$ & $5.96\pm0.35$ & $2.70\pm0.27$\\
   n00580b & 580.7\dotfill & B & $11.97\pm0.48$ & $4.91\pm0.20$ & $4.41^{+1.10}_{-1.65}$ & $6.20\pm0.48$ & $3.13\pm0.29$\\
   n00582b & 582.6\dotfill & B & $12.10\pm0.48$ & $4.49\pm0.18$ & $5.07^{+1.69}_{-1.27}$ & $5.69\pm0.43$ & $2.80\pm0.21$\\
   n00597b & 597.7\dotfill & B & $12.57\pm0.50$ & $4.24\pm0.17$ & $4.27^{+3.42}_{-2.57}$ & $4.76\pm0.09$ & $1.60\pm0.10$\\
   n00602b & 602.6\dotfill & B & $11.62\pm0.47$ & $4.61\pm0.18$ & $4.35^{+2.61}_{-2.18}$ & $6.07\pm0.34$ & $2.79\pm0.21$\\
   n00604b & 604.6\dotfill & B & $12.00\pm0.48$ & $4.50\pm0.18$ & $5.10^{+1.70}_{-1.70}$ & $5.22\pm0.14$ & $2.02\pm0.22$\\
   n00608b & 608.6\dotfill & B & $11.96\pm0.48$ & $4.16\pm0.17$ & $3.98^{+1.59}_{-1.19}$ & $4.81\pm0.12$ & $2.04\pm0.22$\\
   n00610b & 610.6\dotfill & B & $10.99\pm0.44$ & $4.24\pm0.17$ & $5.99^{+1.84}_{-1.38}$ & $5.10\pm0.11$ & $1.99\pm0.20$\\
   n00627b & 627.6\dotfill & B & $11.26\pm0.45$ & $4.01\pm0.16$ & $5.66^{+3.40}_{-1.70}$ & $4.77\pm0.38$ & $1.97\pm0.27$\\
   n00630b & 630.7\dotfill & B & $11.14\pm0.45$ & $3.88\pm0.16$ & $4.35^{+1.74}_{-2.61}$ & $5.08\pm0.31$ & $2.60\pm0.19$\\
   n00635b & 635.6\dotfill & B & $11.87\pm0.47$ & $4.24\pm0.17$ & $5.43^{+3.62}_{-1.81}$ & $5.48\pm0.66$ & $2.77\pm0.33$\\ 
   n00641b & 641.6\dotfill & B & $12.33\pm0.49$ & $4.71\pm0.19$ & $6.36^{+1.82}_{-2.73}$ & $6.22\pm0.32$ & $3.30\pm0.16$\\ 
   n00643b & 643.6\dotfill & B & $10.75\pm0.43$ & $4.46\pm0.18$ & $4.95^{+2.83}_{-2.12}$ & $5.66\pm0.18$ & $2.66\pm0.13$\\
   n00775a & 775.0\dotfill & A & $13.29\pm0.27$ & $5.04\pm0.10$ & $6.29^{+1.93}_{-2.41}$ & $6.82\pm0.19$ & $3.11\pm0.11$\\
   n00777b & 777.0\dotfill & B & $12.69\pm0.51$ & $5.18\pm0.21$ & $6.53^{+1.87}_{-2.34}$ & $6.68\pm0.36$ & $3.11\pm0.14$\\ 
   n00782b & 782.0\dotfill & B & $12.86\pm0.51$ & $5.41\pm0.22$ & $7.21^{+2.06}_{-1.55}$ & $6.78\pm0.25$ & $2.96\pm0.13$\\
   n00787b & 787.0\dotfill & B & $13.49\pm0.54$ & $5.27\pm0.21$ & $7.39^{+3.17}_{-1.59}$ & $6.83\pm0.22$ & $3.83\pm0.12$\\
   n00810b & 810.1\dotfill & B & $14.06\pm0.56$ & $3.95\pm0.16$ & $7.11^{+3.16}_{-1.19}$ & $5.12\pm0.34$ & $2.22\pm0.19$\\
   n00834b & 834.0\dotfill & B & $12.00\pm0.48$ & $4.85\pm0.19$ & $8.14^{+4.79}_{-1.92}$ & $5.65\pm0.20$ & $1.62\pm0.32$\\
   n00838b & 838.1\dotfill & B & $11.44\pm0.46$ & $4.59\pm0.18$ & $5.93^{+2.74}_{-3.20}$ & $5.78\pm0.27$ & $2.23\pm0.13$\\
   n00839a & 839.0\dotfill & A & $11.89\pm0.24$ & $4.89\pm0.10$ & $5.60^{+1.29}_{-1.29}$ & $6.31\pm0.25$ & $2.85\pm0.15$\\
   n00840b & 840.1\dotfill & B & $12.50\pm0.50$ & $4.91\pm0.20$ & $6.77^{+2.08}_{-1.56}$ & $5.85\pm0.27$ & $2.01\pm0.10$\\
   n00843b & 843.0\dotfill & B & $12.45\pm0.50$ & $4.87\pm0.19$ & $5.74^{+1.91}_{-2.78}$ & $5.55\pm0.26$ & $1.62\pm0.14$\\
   n00847b & 847.0\dotfill & B & $12.33\pm0.49$ & $4.82\pm0.19$ & $7.12^{+2.85}_{-1.90}$ & $6.10\pm0.34$ & $2.14\pm0.14$\\
   n00867b & 867.0\dotfill & B & $12.25\pm0.49$ & $4.20\pm0.17$ & $6.53^{+2.18}_{-1.09}$ & $4.95\pm0.59$ & $1.31\pm0.13$\\
   n00869b & 869.0\dotfill & B & $12.55\pm0.50$ & $3.85\pm0.15$ & $4.95^{+1.98}_{-1.98}$ & $4.79\pm0.27$ & $1.71\pm0.13$\\
   n00872b & 873.0\dotfill & B & $12.15\pm0.49$ & $4.09\pm0.16$ & $4.74^{+2.11}_{-2.64}$ & $4.88\pm0.27$ & $1.93\pm0.10$\\
   n00876a & 876.0\dotfill & A & $12.58\pm0.50$ & $4.31\pm0.17$ & $5.27^{+2.39}_{-2.39}$ & $5.78\pm0.20$ & $3.31\pm0.10$\\
   n00877b & 877.0\dotfill & B & $12.44\pm0.50$ & $4.37\pm0.17$ & $6.11^{+2.22}_{-1.67}$ & $5.54\pm0.21$ & $2.13\pm0.11$\\
   n00886a & 885.8\dotfill & A & $11.54\pm0.23$ & $4.18\pm0.08$ & $4.88^{+1.95}_{-1.95}$ & $5.43\pm0.20$ & $2.71\pm0.12$\\
   n00892b & 892.8\dotfill & B & $12.14\pm0.49$ & $4.54\pm0.18$ & $5.83^{+1.94}_{-1.94}$ & $5.42\pm0.23$ & $1.85\pm0.11$\\
   n00906b & 906.9\dotfill & B & $11.44\pm0.46$ & $4.53\pm0.18$ & $5.64^{+2.82}_{-1.88}$ & $5.73\pm0.27$ & $2.22\pm0.14$\\
   n00921b & 921.9\dotfill & B & $11.61\pm0.46$ & $4.53\pm0.18$ & $6.35^{+2.93}_{-2.93}$ & $5.57\pm0.28$ & $1.91\pm0.12$\\
   n00927a & 926.8\dotfill & A & $11.13\pm0.22$ & $4.43\pm0.09$ & $4.62^{+1.85}_{-1.85}$ & $5.81\pm0.17$ & $2.57\pm0.14$\\
   n00930b & 930.8\dotfill & B & $11.43\pm0.46$ & $4.01\pm0.16$ & $4.89^{+2.67}_{-1.87}$ & $5.00\pm0.34$ & $1.74\pm0.15$\\
   n00935b & 935.9\dotfill & B & $11.92\pm0.48$ & $4.51\pm0.18$ & $5.35^{+1.89}_{-1.89}$ & $5.68\pm0.40$ & $2.85\pm0.20$\\
   n00936b & 936.6\dotfill & B & $11.23\pm0.45$ & $4.39\pm0.18$ & $5.30^{+1.77}_{-2.66}$ & $5.30\pm0.27$ & $1.84\pm0.11$\\
   n00949b & 949.6\dotfill & B & $12.03\pm0.48$ & $5.04\pm0.20$ & $5.20^{+1.49}_{-2.24}$ & $6.06\pm0.19$ & $2.15\pm0.09$\\
   n00961b & 961.6\dotfill & B & $11.32\pm0.45$ & $4.90\pm0.20$ & $6.29^{+1.94}_{-1.46}$ & $5.94\pm0.31$ & $2.32\pm0.11$\\
   n00966b & 966.7\dotfill & B & $11.60\pm0.46$ & $4.39\pm0.18$ & $4.47^{+2.24}_{-2.69}$ & $5.23\pm0.24$ & $2.09\pm0.12$\\
   n00981a & 980.7\dotfill & A & $11.72\pm0.21$ & $4.73\pm0.08$ & $5.88^{+2.94}_{-1.47}$ & $6.34\pm0.06$ & $2.79\pm0.04$\\
   n00983b & 983.7\dotfill & B & $11.41\pm0.46$ & $4.46\pm0.18$ & $5.88^{+2.20}_{-1.83}$ & $5.37\pm0.32$ & $2.00\pm0.14$\\
   n00988b & 988.6\dotfill & B & $11.56\pm0.46$ & $4.30\pm0.17$ & $5.26^{+2.39}_{-2.39}$ & $5.05\pm0.27$ & $1.72\pm0.10$\\
   n00991b & 991.7\dotfill & B & $11.35\pm0.45$ & $4.15\pm0.17$ & $6.69^{+2.09}_{-1.25}$ & $5.07\pm0.24$ & $1.92\pm0.12$\\
   n00994b & 994.6\dotfill & B & $11.37\pm0.46$ & $4.11\pm0.16$ & $6.48^{+2.99}_{-1.50}$ & $5.02\pm0.26$ & $1.49\pm0.13$\\
   n01009b & 1009.6\dotfill& B & $10.98\pm0.44$ & $3.76\pm0.15$ & $4.50^{+1.80}_{-2.25}$ & $4.50\pm0.30$ & $1.55\pm0.13$\\
   n01012b & 1012.6\dotfill& B & $12.18\pm0.49$ & $3.70\pm0.15$ & $5.71^{+2.59}_{-1.55}$ & $4.57\pm0.21$ & $1.79\pm0.12$\\
   n01019b & 1019.6\dotfill& B & $12.75\pm0.51$ & $4.20\pm0.17$ & $5.89^{+3.63}_{-2.72}$ & $5.27\pm0.33$ & $2.80\pm0.15$\\
   n01022b & 1022.6\dotfill& B & $12.09\pm0.48$ & $4.57\pm0.18$ & $7.26^{+3.63}_{-3.63}$ & $5.52\pm0.40$ & $1.58\pm0.16$\\

   \end{longtable}  
  
     \begin{list}{}{}
   \item[$^\mathrm{a}$]The Julian Date is scaled to zero point at JD = 2,450,000.
   \item[$^\mathrm{b}$]In units of $\rm{10^{-15}\ ergs\ s^{-1}\ cm^{-2}\ \AA^{-1}}$. The values are obtained by Peterson
                       et al.(2000).
   \item[$^\mathrm{c}$]In units of $\rm{10^{-13}\ ergs\ s^{-1}\ cm^{-2}}$. The fluxes are obtained from Peterson et al.(2000).
   \item[$^\mathrm{d}$]In units of $\rm{10^{-13}\ ergs\ s^{-1}\ cm^{-2}}$.
   \item[$^\mathrm{e}$]Each flux includes the contributions of all three components.
   \item[$^\mathrm{f}$]Each flux contains the contributions of a broad base and a narrow peak.
   \end{list}
 
   \begin{table}
   \centering
   \caption[]{Variations of the \ion{Fe}{ii} complex in six selected Seyfert galaxies}
   \label{FeII}

   $$
   \begin{array}{ccccccc}
   \hline
   \noalign{\smallskip}
   \rm{Source\ name} && \rm{FWHM}^{\mathrm{a}} && \rm{R_{Fe}\ vs.\ F_{\lambda}}^{\mathrm{b}}&& \rm{Reference}\\
   &&\rm{(km\ s^{-1})}&&&&\\
   \hline
   \hline
   \noalign{\smallskip}
   \rm{NGC\,7603}\dotfill && \rm{6560}  &&  \rm{Negative} && 1\\
   \rm{Mark\,110}\dotfill && \rm{1515}  &&  \rm{Negative} && 2\\
   \rm{NGC\,4051}\dotfill && \rm{1100}  &&  \rm{Positive} && ....\\
   \rm{Mark\,359}\dotfill && \rm{900}   &&  \rm{Positive} && 3,4 \\
   \rm{Mark 1044}\dotfill && \rm{1010}  &&  \rm{Positive} && 3,4\\
   \rm{Akn\,564}\dotfill  && \rm{865}   &&  \rm{Positive} && 3,4\\
   \hline
   \noalign{\smallskip}
   \end{array}
   $$
   \begin{list}{}{}
    \item[$^\mathrm{a}$] Each of the FWHM measured from the mean profile of the H$\beta$ line.
    \item[$^\mathrm{b}$] The relation between $\rm{R_{Fe}}$ and continuum. Positive: the relation is positive; 
    Negative: the relation is negative. 
    \item[Notes] 1: Kollatschny et al. (2000); 2: Kollatschny et al. (2001); 3: Giannuzzo \& Stirpe (1996);
     4: V\'{e}ron-Cetty et al. (2001)
   \end{list}

   \end{table}

\end{document}